\documentstyle[preprint,aps,prb,epsfig]{revtex}
\begin{document}
\draft
\tightenlines

\title{Magnetoplasmon excitations in an array of periodically
modulated quantum wires}

\author{B. P.~van Zyl} 
\address{Institute for Microstructural Sciences, National Research Council of
Canada, Ottawa, Ontario, K1A 0R6}
\author{E.~Zaremba}
\address{Department of Physics, Queen's University, Kingston, 
Ontario, Canada K7L 3N6}

\maketitle
\begin{abstract}
Motivated by the recent experiment of Hochgr\"afe {\em et al.,} we have
investigated the magnetoplasmon excitations in a periodic array of
quantum wires with a periodic modulation along the wire direction.
The equilibrium and dynamic properties of the system are
treated self-consistently within the
Thomas-Fermi-Dirac-von Weizs\"acker approximation.
A calculation of the dynamical response of the system to a far-infrared 
radiation field reveals a resonant anticrossing between the Kohn mode 
and a finite-wavevector longitudinal excitation which is
induced by the density modulation along the wires. 
Our theoretical calculations are found to be in excellent agreement with
experiment. 
\end{abstract}

\section{INTRODUCTION}
\label{introduction}

Modern microstructuring techniques have made it possible to 
experimentally study a wide variety of low-dimensional electron systems
such as quantum dots, wires and rings.\cite{kirk-davies}
These systems are usually fabricated starting from a two-dimensional 
electron gas (2DEG) located at a modulation-doped semiconductor 
heterojunction.  The application of laterally microstructured metallic 
gates\cite{mackens1984} or the use of deep-mesa etching 
techniques\cite{heitmann1989} allow for the precise patterning of the 
2DEG.  Over the past 15 years, a great deal of information has been
obtained about the transport and far-infrared (FIR) optical
properties of these systems. For example, magneto-transport measurements
of laterally modulated 2DEG's reveal interesting magnetoresistance 
oscillations associated with the commensurability of the cyclotron 
radius and modulation period $a$.\cite{kirk-davies}  
Far-infrared transmission experiments, on the other hand, reveal
a rich spectrum of collective excitations whose frequencies
exhibit novel magnetic field dispersions.\cite{heitmann1993}

In this paper, we are interested in the magnetoplasmon excitations of an
array of parallel quantum wires which are also periodically 
modulated along their length. Such a system was recently fabricated by
Hochgr\"afe {\em et al.,}\cite{hochgraefe} and studied using FIR
transmission spectroscopy. They observe a pair of collective
excitations which appear to interact and anticross as a function
of magnetic field. This behaviour is qualitatively explained by 
elementary considerations of the magnetoplasmon mode spectrum of a
uniform wire as described by the theory of Eliasson 
{\em et al.}\cite{eliasson1986a} The periodic modulation along the
length of the wires acts as a grating-coupler which couples the Kohn
mode to a finite-wavevector 1D magnetoplasmon of the uniform wire. 
The experimental data also show a novel kink-like feature in the 
vicinity of the cyclotron frequency $\omega_c$, that was neither
discussed, nor explained in Ref.~[5].

Our purpose here is to theoretically investigate the system studied
by Hochgr\"afe {\em et al.} in order to better understand the nature of
the magnetoplasmon excitations observed. In particular, we would like to
gain some insight into the nature of the periodic modulation of the 
electronic density in these systems, and how this leads to the coupling 
of the modes discussed above.
To this end, we adopt a hydrodynamic theory based on the 
Thomas-Fermi-Dirac-von Weizs\"acker (TFDW) approximation. This approach
has already been used to describe the collective excitations in 
laterally modulated 2DEG's~\cite{vanzyl1999a} 
as well as other low-dimensional electronic 
structures.\cite{zaremba1994,zaremba1996,vanzyl1999b,vanzyl1999c} Its
main virtue is that it treats the ground state and dynamic properties 
of the system in a consistent fashion.~\cite{footnote1}
Our earlier successful applications give us confidence that the
approach is able to accurately describe the collective
excitations in the systems studied in Ref.~[5].  

The plan of the paper is as follows.  In Sec. \ref{modulated}
we define our model for a periodically modulated quantum wire array.
Then, we systematically analyze the effects of this modulation on the 
ground state, and dynamical properties of the system.
To the extent possible, we make direct comparisons with the
experiment of Ref.~[5].
Finally, in Sec. \ref{conclusions}, we present our concluding remarks.

\section{MODULATED WIRES}
\label{modulated}

The TFDW approximation to the ground state of a quantum confined 2DEG is
given by (atomic units are used throughout) 
\cite{vanzyl1999a,vanzyl1999b,vanzyl1999c}

\begin{eqnarray}
E[n] &=& \int d^2{\bf r} \left[ \frac{\pi}{2} n^2 + \frac{\lambda_w}{8}
\frac{\mid \nabla n(\bf{r}) \mid^{2}}{n(\bf{r})} -
\frac{4}{3}\sqrt{\frac{2}{\pi}} n^{3/2} \right] \nonumber \\ &+&
\frac{1}{2} \int d^2{\bf r} \int d^2{\bf r}'
\frac{n({\bf r}) n({\bf r}')}{\mid {\bf r} - {\bf r}' \mid} +
\int d^2{\bf r}\, v_{\rm ext}({\bf r}) n({\bf r})\,.
\label{functional}
\end{eqnarray}
The first term is the Thomas-Fermi approximation to the kinetic energy,
the second term is a von Weizs\"acker-like gradient correction, and
the third term is the Dirac local exchange energy. For simplicity we
neglect a correlation energy contribution which in any case has a small
effect. The last two terms characterize the Hartree self-energy of the 
electrons and the interaction with the external potential, respectively.
Once the confining potential $v_{\rm ext}({\bf r})$ has been
defined, the variational minimum of Eq.~(\ref{functional}) yields the
ground state electronic density, $n_0({\bf r})$, for the system of 
interest.~\cite{vanzyl1999a,vanzyl1999b,vanzyl1999c}
To describe the array of modulated quantum wires in 
[5] we adopt a cofining potential having the 
form~\cite{justify}

\begin{equation}
v_{\rm ext}(x,y) = \frac{1}{2} \omega_0^2 x^2 + V_y{\rm cos}(Gy)~,
\label{v_ext}
\end{equation}
with $G=2\pi/a$.  
It is understood that the potential in Eq.~(\ref{v_ext}) is
centered in each unit cell and periodically extended throughout the
array with period $a$.  Setting $V_y=0$ describes a periodic array of 
parabolically confined quantum wires, whereas $V_y \neq 0$ 
provides a modulation along the wire direction, also with period $a$ 
(see Fig.~\ref{fig1}).
Keeping in mind that the samples in [5]
are prepared from a modulation-doped AlGaAs/GaAs heterostructure,
the appropriate physical parameters are:
the effective Bohr radius $a_0^{\star} = 103$ {\AA} and the effective
Rydberg of energy, ${\rm Ry}^{\star} = e^2/2\epsilon a_0^{\star}=$
5.4 meV.  In these units, a 2D electronic density of
$1.0\times 10^{11} {\rm cm}^{-2}$ is given by
$\overline{n}_{\rm 2D} = 0.10 (a_0^{\star})^{-2}$,
and an energy of $0.5~{\rm a.u} = 1{\rm Ry}^{\star}$ is equivalent to
43.57 cm$^{-1}$.

In Fig.~\ref{fig1} we present the equilibrium density distributions for
a small section of the array. 
Figure~\ref{fig1}(a) is generated with
the following physical parameters: an average density 
$\overline {n}_{\rm 2D}=1.0 \times 10^{11}~{\rm cm}^{-2}$,
$a = 600~{\rm nm}$,
$\omega_0 = 0.427$ a.u, and $V_y = 1$ a.u. 
As will be discussed in more
detail later, this choice allows us to reproduce much of the
experimental results~\cite{hochgraefe} obtained with the sample
having a density of $\overline {n}_{\rm 2D}=1.2
\times 10^{11}~{\rm cm}^{-2}= 0.12 (a_0^{\star})^{-2}$.
We therefore believe that the very strongly modulated density shown is
typical of the densities studied in the experiments. In
Fig.~\ref{fig1}(b), we show the effect of keeping the confining
potential fixed and doubling the 2D density to
$\overline {n}_{\rm 2D}=0.2$. It can be seen that the degree of
corrugation is reduced in both directions, and that the densities from
adjacent wires are now beginning to
overlap.  In going from
Fig.~\ref{fig1}(b)
to Fig.~\ref{fig1}(c) we have doubled the strength of the modulation to
$V_y=2$ a.u. 
This gives rise to a much stronger modulation of the density along the 
length of the wires which now appear as a string of quantum dots. This 
is the density profile used to simulate the FIR response of the system
as described in the second half of the experimental study in
Ref.~[5].  Finally, in Fig.~\ref{fig1}(d), we
show the density profile generated with the same potential parameters
as in Fig.~\ref{fig1}(c), but with a further $50\%$ increase in the 
average density. This density takes on the character of an array of 
coupled quantum dots.

The dynamics of a mesoscopic system within the TFDW approach is 
governed by a set of (linearized) hydrodynamic equations, 
\cite{vanzyl1999a,vanzyl1999b,vanzyl1999c}

\begin{equation}
\frac{\partial \delta n}{\partial t}+\nabla \cdot (n_{0} {\bf v}) = 0~,
\label{lincont}
\end{equation}
with $\delta n({\bf r},t) = n({\bf r},t) - n_0({\bf r})$, and
\begin{equation}
\frac{\partial {\bf v}}{\partial t} = -\gamma {\bf v} 
-\nabla U^{\rm int} + {\bf F}^{\rm ext}~.
\label{linmom}
\end{equation}
Eq.~(\ref{lincont}) is the usual continuity equation, and
Eq.~(\ref{linmom}) is a statement of Newton's second law. The quantity 
$U^{\rm int}$ plays the role of an effective potential for the internal 
force fluctuations, and ${\bf F}^{\rm ext}$ is the force due to any
additional external electromagnetic fields imposed on the system,
including a homogeneous, static magnetic field oriented normal to the 
2DEG. In Eq.~(\ref{linmom}), we
have also included a phenomenological viscous damping term, 
$-\gamma {\bf v}$, to account for Drude scattering.  This term has the
effect of giving the collective modes a finite lifetime.

The solution of Eqs.~(\ref{lincont}) and (\ref{linmom}) allows us to 
calculate the physically relevant, time-averaged FIR power 
absorption.~\cite{vanzyl1999a,vanzyl1999b,vanzyl1999c,footnote2} 
In terms of the induced current density
${\bf j}^{\rm ind}({\bf r},t) = -n_0({\bf r}) {\bf v}({\bf r},t)$, 
we have

\begin{eqnarray}
\langle P \rangle_{t} &=&
\left \langle \int d^2{\bf r}~{\bf j}^{\rm ind}({\bf r},t)\cdot
{\bf E}^{\rm ext}({\bf r},t) \right \rangle_{t}~, 
\label{avgpwr}
\end{eqnarray}
where 
\begin{equation}
{\bf E}^{\rm ext}({\bf r},t) = 
\left(E_{x}\hat{\bf x} + E_{y}\hat{\bf y}\right)
{\rm cos}(\omega t)
\end{equation}
is a spatially uniform radiation field polarized along ($E_{x}=0$) 
or perpendicular to ($E_{y}=0$) the direction of the wires.
Most of our calculations are carried out for
the perpendicular polarization, but some calculations are also done 
for the parallel polarization in order to identify some of the 
longitudinal excitations at low magnetic fields ($B \lesssim 0.5$T).

The theoretical FIR power absorption for the density profile 
corresponding to Fig.~\ref{fig1}(a) is shown in 
Fig.~\ref{fig2}.~\cite{footnote3} 
The curves are displayed in steps of 
$\Delta B=0.25$T starting at $B=0$T, and are offset vertically for clarity.
The dashed lines are guides to the eye and indicate the peak trajectories;
the open symbols at the end of each line serve to label the different
magnetoplasmon modes (see Fig.~\ref{fig3}).
The $B=0$T trace shows an isolated peak at
$\omega \simeq 37~{\rm cm}^{-1}$ which is very close to the
parabolic confinement frequency $\omega_0$.
This is a clear signature of the center-of-mass Kohn mode (KM)
predicted by the generalized Kohn theorem,~\cite{dobson94} and 
corresponds to a rigid oscillation of the equilibrium density
perpendicular to the wires.
With increasing magnetic field, this peak at first shifts slightly to
higher frequencies, but around $B \simeq 0.8$T, this trend is reversed
with the appearance of a small shoulder to the high frequency side of
the peak. The shoulder evolves into a separate peak which gains in
oscillator strength and shifts to higher frequencies along the dashed
line ending with the unfilled diamond. Although difficult to
resolve, there is also a weak peak labeled by the unfilled triangle.

The positions of the absorption peaks as a function of magnetic field
are displayed in Fig.~\ref{fig3}. The unfilled symbols used in this
figure correspond to the labels used to track the peak trajectories
in Fig.~\ref{fig2}. For comparison, the filled squares are the
raw experimental data taken from Ref.~[5].
The dashed lines in Fig.~\ref{fig3} are the cyclotron frequency
$\omega_c$ and $2\omega_c$.  We will discuss the
solid curves below.
The splitting of the upper experimental mode between $B=1.5$T and 2.5T
is attributed in [5] to the presence of the Bernstein mode
at $\omega = 2
\omega_c$.~\cite{demel1991,drexler92,gudmundsson95}
Apart from this detail, it is clear that the TFDW results provide a
smooth interpolation through the experimental points, and overall are
in excellent agreement with the data.  In
particular, the size
and position of the anticrossing between the two lowest-lying branches
(open circles and open diamonds) is well reproduced.
 
In order to explain the details of the magnetic dispersions in
Fig.~\ref{fig3},
it will prove useful to first consider the simpler case of a
2DEG which is modulated in one direction (see Ref.~[7] for details).
These systems were generated by setting $\omega_0 = 0$ in 
Eq.~(\ref{v_ext}) and taking the
modulation to be in the $x$-direction, i.e., $V_y \rightarrow V_x$.
An additional weak density modulation along the wires will induce
a coupling between the $q_y=0$ modes of the wire to modes with 
wavevector $q_y \equiv G = 2\pi/a$.
It is therefore of interest to see how the $q_y=2\pi/a$ mode frequencies
evolve as a function of magnetic field. 
In Figs.~\ref{fig4}(a)-(c), we present the magnetic dispersions of a
uniaxially modulated 2DEG at fixed $q_y = 2\pi/a$.
The excitation spectrum for the weakly modulated system ($V_x = 0.25$ a.\,u) 
is shown in Fig.~\ref{fig4}(a) and essentially exhibits the magnetoplasmon 
dispersion of a homogeneous 2DEG.  The dense spectrum of
modes starting from $\omega=0$ at $B=0$T are a result of the broken
translational invariance in the modulation direction, and eventually
evolve into a series of edge magnetoplasmons (EMP's).
In Fig.~\ref{fig4}(b) we show the magneto-dispersion for the moderately
modulated 2DEG having a modulation $V_x = 1.25$ a.\,u.
For this strength of modulation, one of the magnetoplasmon branches 
starting from $\omega \simeq 36$ cm$^{-1}$ acquires a weak negative 
dispersion and exhibits an anticrossing-like behaviour with a mode which 
has the character of an EMP.\cite{fetter}
The mode dispersing upward from $\omega \simeq 45$ cm$^{-1}$ is
associated with the KM and continues to increase in frequency with
increasing magnetic field.~\cite{fetter} Finally, 
in Fig.~\ref{fig4}(c) we show the spectrum of the $q_y = 2\pi/a$ 
modes for a modulation amplitude of $V_x = 1.5$ a.\,u.  
In this modulation regime, the 1DMP has a pronounced
negative dispersion for $B \lesssim 2~T$, and the gap
associated with the anticrossing between the $q_y = 2\pi/a$ 1DMP 
and EMP modes has decreased.

We are now in a better position to comment on the details of the magnetic
dispersions in Fig.~\ref{fig3}. Several solid curves are shown which
correspond to the smooth wire limit. The solid curve which is lowest at
$B=0$ is the dispersion relation of the $q_y=0$ center-of-mass KM, 
namely,~\cite{vanzyl1999a,bangert1993}
\begin{equation}
\omega^2(B) = \omega_0^2 + \omega_c^2,
\label{KM}
\end{equation}
with $\omega_0 = 37.2~{\rm cm}^{-1}$.  The next two solid curves are
the magnetic dispersions of the two $q_y=2\pi/a$ modes calculated by
taking $V_y=0$ and $\omega_0 = 37.2$ cm$^{-1}$ in Eq. (\ref{v_ext}) and
are analogous to the pair of modes starting from $\omega \simeq 45$
cm$^{-1}$ in Fig.~4(c).  We thus conclude that the two
branches denoted by open triangles and diamonds in Fig.~\ref{fig3} are
derived from the $q_y=2\pi/a$ KM and the $q_y=2\pi/a$ 1DMP of the 
uniform wire array as discussed above.  However, in the present
situation, the $q_y=2\pi/a$ 1DMP is in fact coupled to the $q_y = 0$ KM 
by the periodic modulation along the wires, which results in the 
anticrossing behaviour seen in
the vicinity of $B=1$ T. For $B \lesssim 1$T, the mode indicated by open
circles has the character of the KM, while for $B \gtrsim 1$T, the mode
changes over into the $q_y=2\pi/a$ 1D EMP. A similar change in character
occurs for the open diamond branch, but in the opposite sense. This is
precisely the interpretation of the experimental results given in
Ref.~[5]. The highest energy branch (open
triangles) corresponding to the weak high energy resonance in
Fig.~\ref{fig3} is clearly the $q_y = 2\pi/a$ mode belonging to the
upward dispersing bulk mode (see e.g., Fig.~\ref{fig4}(c)).
Finally, we note that the track of the lowest mode in
Fig.~\ref{fig3} shows a small `kink' at $\omega = \omega_c$.
As we shall see below, this kink becomes more prominent as the density
of the system is increased. We defer further discussion of this feature
until then.

We next consider the effect of increasing the electronic density. This
is achieved experimentally by means of illumination, and in one
particular run resulted in an approximate doubling of the average
density.  To simulate this situation, we considered
a system having a density of $\overline{n}_{\rm 2D} = 0.2$
and the same confining potential used to generate the results in
Fig.~\ref{fig3}. The density distribution in this case is shown in
Fig.~\ref{fig1}(b) where it can be seen that the modulation of the
density along the wires is weaker than for $\overline{n}_{\rm 2D} =
0.1$.  However, the simple assumption of an increase in the electron
density was not able to account for the observed trends in the
experimental data. Indeed, agreement with experiment could only be
achieved by simultaneously increasing the amplitude of the modulation
potential to $V_y = 2$ a.\,u. The corresponding density distribution
is shown in Fig.~\ref{fig1}(c). The modulation of the density along the
wires is clearly more pronounced than in Figs.~\ref{fig1}(a) or
\ref{fig1}(b). In Fig.~\ref{fig5}, we present the calculated FIR
power absorption.
As in Fig.~\ref{fig2}, there is a
strong KM peak at $B=0$T, consistent with the parabolic confinement
in the $x$-direction. In addition, due to the increased coupling of the
density in a direction perpendicular to the wires, we pick up some
weak Drude absorption at $\omega=0$. The dashed curve labeled by the
open square symbol tracks the location of this peak which follows
closely the dispersion of the cyclotron mode $\omega = \omega_c$ up to
the point where it interacts weakly with the 1DMP (open circles).
 
The peak positions as a function of $B$ are shown in Fig.~\ref{fig6}.
As compared to Fig.~\ref{fig3}, the anticrossing between the KM and the
1DMP has now been pushed down in magnetic field so that there is already a
strong interaction at $B=0$ T. As a result, the frequency of the lower
mode indicated by open circles decreases monotonically for the entire
range of magnetic fields presented, and at higher magnetic fields has
the character of the 1D EMP. The mode indicated by open diamonds
attains a noticeable signal at $B \approx 0.8$T as it evolves into
the KM. However, unlike the case of lower electron density, the
oscillator strength
of the KM is shared with higher lying modes indicated by the open
triangles and stars in Fig.~\ref{fig6}. Although the triangle mode in
the figure is very weak, it appears more prominently in the FIR 
absorption spectrum when the radiation is polarized along the wires. 
This mode is the same $q_y = 2\pi/a$ mode discussed earlier (triangles 
in Fig.~\ref{fig3}) but in this case it interacts with another mode
near $B=1.5$T and
continues along the starred branch at higher magnetic fields where its
strength is comparable to that of the KM. 

The agreement with the experimental data (filled squares) is again
very good, except in the vicinity of the Bernstein mode at $2\omega_c$.
Our model provides a smooth and continuous interpolation through this
region. The lower experimental branch closely follows the 1DMP
indicated by open circles while the upper branch is clearly associated
with the KM (open diamonds).
 
The weak structure in the trajectory of the lower experimental branch
seen near $\omega_c$ can be explained in terms of the pair of
$q_y=2\pi/a$ modes which appear to anticross in Fig.~4(c). In the
present situation, these modes are excited by virtue of the $2\pi/a$
modulation along the wire length.  To confirm 
our interpretation of this kink-like feature, we considered a further
increase in the density of the system without changing the potential
used to generate Fig.~\ref{fig1}(c). The density profile associated
with this situation is shown in Fig.~\ref{fig1}(d). Figure \ref{fig7}
illustrates the FIR power absorption of the system, and
Fig.~\ref{fig8} shows the peak positions as a function of the magnetic
field. 
Comparing Fig.~\ref{fig8} with Fig.~\ref{fig4}(c), we see that these 
dispersions share several common characteristics.  In particular,
we see that the open diamond and triangles in Fig.~\ref{fig8} have the
same qualitative behaviour as the two $q_y=2\pi/a$ modes
starting at $\omega \simeq 45~{\rm cm}^{-1}$
in Fig.~\ref{fig4}(c).  Also, the strong anticrossing between the
open circles and open squares in Fig.~\ref{fig8} clearly has the same origin
as the anticrossing between the 1DMP and the EMP mode
in Fig.~\ref{fig4}(c).
Thus, the weak ``kink'' in the experimental data of
Ref.~[5] is associated with the anticrossing of the $q_y = 2\pi/a$
1DMP and EMP modes which are excited by the
$G=2\pi/a$ modulation along the wire direction.
As Fig.~4 indicates, the apparent coupling between the 1DMP and EMP
modes decreases as the magnitude of the uniaxial modulation increases.
This explains why the ``kink'' feature is more apparent at higher 
densities and gets progressively weaker as the wires become
increasingly isolated from each other.

\section{CONCLUSIONS}	
\label{conclusions}

In this paper, we have studied the magnetoplasmon excitations in an array of
periodically modulated quantum wires within the TFDW approximation.  
We have shown that the modulation along the wire direction 
effectively acts as a grating coupler which induces a coupling 
between the ${\bf q}=0$ Kohn mode and the
${\bf q}=(0,2\pi/a$) 1D plasmon, resulting in a resonant anticrossing
of these modes at a finite magnetic field.  
We have compared our theoretical results with the data of
Ref.~[5], and found excellent agreement for all density and magnetic 
field regimes investigated.  In addition, we have provided a simple 
explanation for the kink feature at $\omega=\omega_c$, which was 
observed, but not explained, in Ref.~[5]. 
Based on the overall agreement between theory and experiment, we believe
that the
underlying ground state densities used in the calculations, as 
illustrated in Fig.~1, are in fact a realistic representation of the 
density profiles produced in the experiments. 

\acknowledgements
We would like to thank M. Hochgr\"afe, R. Krahne, and D. Heitmann for 
useful discussions and for providing access to their experimental 
data. This work was supported by a grant from the Natural Sciences
and Engineering Research Council of Canada.

\begin{figure}
\caption{
Equilibrium density distributions, presented as surface and
contour plots, for:
(a) $\overline{n}_{\rm 2D} = 0.1$, $V_y = 1$ a.u, (b)
$\overline{n}_{\rm 2D} = 0.2$, $V_y = 1$ a.u,
(c) $\overline{n}_{\rm 2D} = 0.2$, $V_y = 2$ a.u,
and (d) $\overline{n}_{\rm 2D} = 0.3$, $V_y = 2$ a.u.
In all cases, $a=600~{\rm nm}$ and $\omega_0 = 0.427$ a.u.
The parameters pertaining to $v_{\rm ext}$ were chosen to reproduce the
$B=0$T excitation frequencies in Ref.~[5].} 
\label{fig1}
\end{figure}
 
\begin{figure}
\caption{
Calculated FIR power absorption of the modulated quantum wire
array corresponding to the density in Fig.~\ref{fig1}(a).  The unfilled
symbols are used to label the various peak trajectories, denoted by the
dashed lines.  The incident radiation is polarized in a direction
perpendicular to the wires.}
\label{fig2}
\end{figure}

\begin{figure}
\caption{
Theoretical $B$ dispersion of the resonant FIR absorption data
in Fig.~\ref{fig2}.  The unfilled symbols correspond to those in 
Fig.~\ref{fig2}, and the filled squares are the 
experimental data from
Ref.~[5].  The dashed and full curves are explained in
the text.}
\label{fig3}
\end{figure}

\begin{figure}
\caption{
Calculated $B$ dispersions of the modes in a uniaxially modulated 
2DEG at a fixed wavevector ${\bf q} = (0,2\pi/a)$. The amplitudes of
the modulating potentials are (a) $V_x = 0.25$
a.\,u.~, (b) $V_x = 1.25$ a.\,u.~, and (c) $V_x = 1.5$ a.\,u.
The dashed line represents the cyclotron frequency.}
\label{fig4}
\end{figure}

\begin{figure}
\caption
{As in Fig.~\ref{fig2} but for an equilibrium density defined by
Fig.~\ref{fig1}(c).}
\label{fig5}
\end{figure}

\begin{figure}
\caption{
As in Fig.~\ref{fig3}, but with the data extracted from Fig.~\ref{fig5}.}
\label{fig6}
\end{figure}

\begin{figure}
\caption{
As in Fig.~\ref{fig2}, but for an equilibrium density defined by
Fig.~\ref{fig1}(d).}
\label{fig7}
\end{figure}

\begin{figure}
\caption{
As in Fig.~\ref{fig3}, but with data extracted from Fig.~\ref{fig7}.}
\label{fig8}
\end{figure}


\begin{references}
\bibitem{footnote1}
That is, the form of the equilibrium density is not chosen arbitrarily.
Perturbations of the system away from the equilibrium state generate
internal forces which drive the system back towards equilibrium.  These
forces are consistently included in the TFDW hydrodynamic equations
used to describe the dynamics of the system.
See,  e.g., Ref.~[7].
\bibitem{footnote2}
Details pertaining to the solution of Eqs. (\ref{lincont}) and
(\ref{linmom}) can be found in Refs.~[7,10].
\bibitem{footnote3}
The FIR power absorption data presented in Figs.~\ref{fig2}, \ref{fig5},
and \ref{fig7}
have been generated with the phenomenological damping parameter $\gamma
= 4.4$ cm$^{-1}$.
However, when peaks lie very close to each other, a smaller value
of $\gamma = 2.2$ cm$^{-1}$ 
allows us to better resolve the peak positions, especially
for peaks having weak resonances.
\end{references}
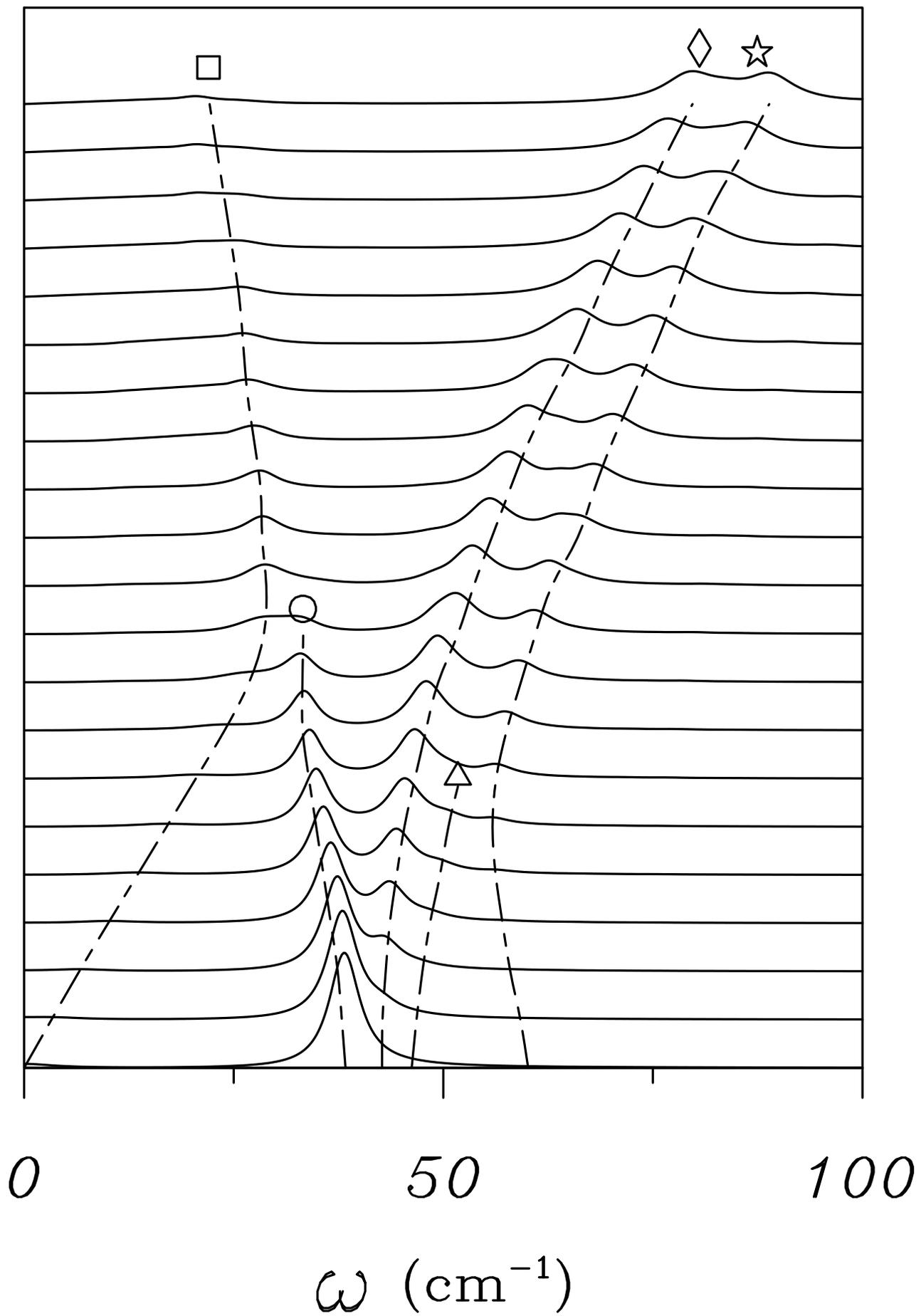
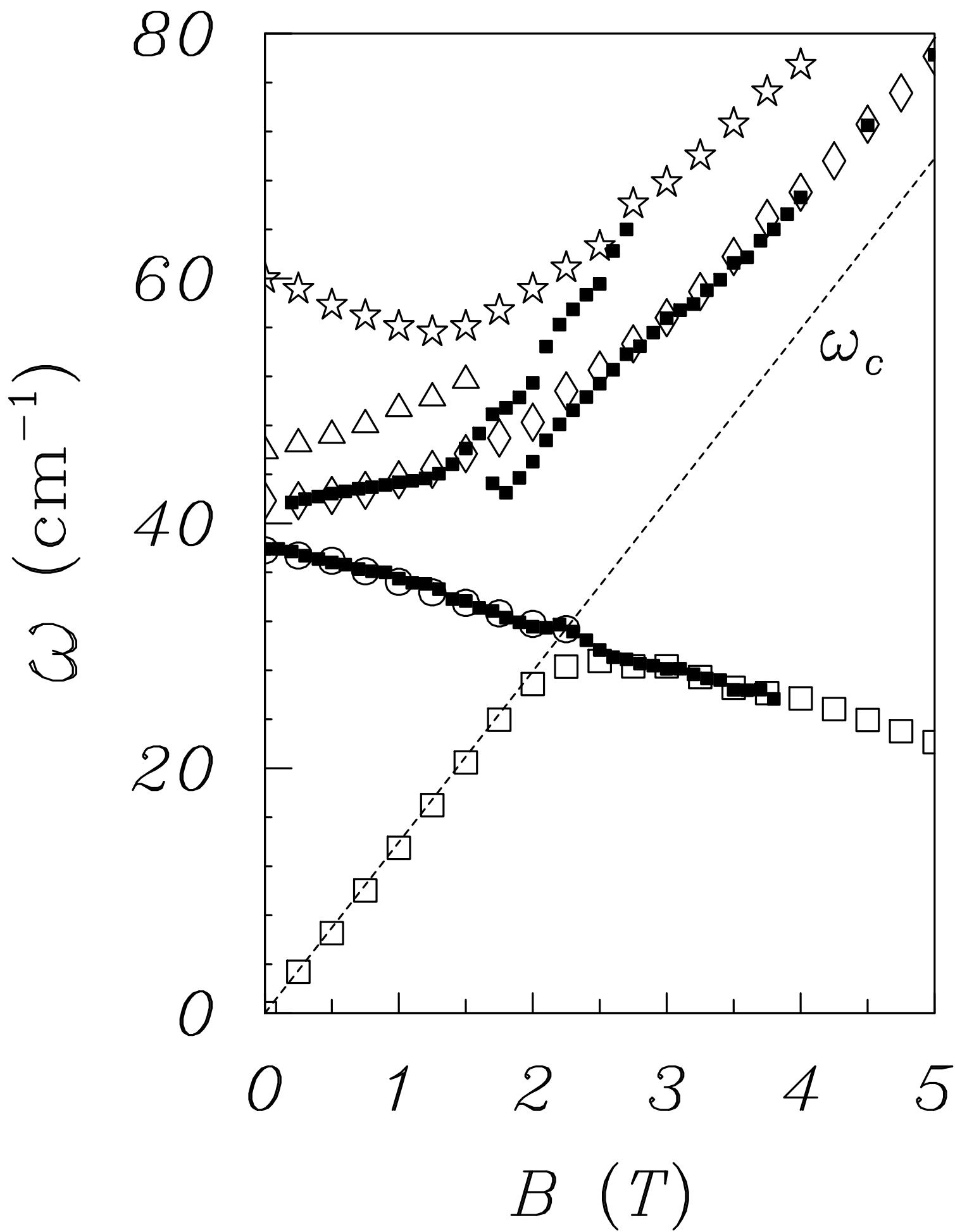
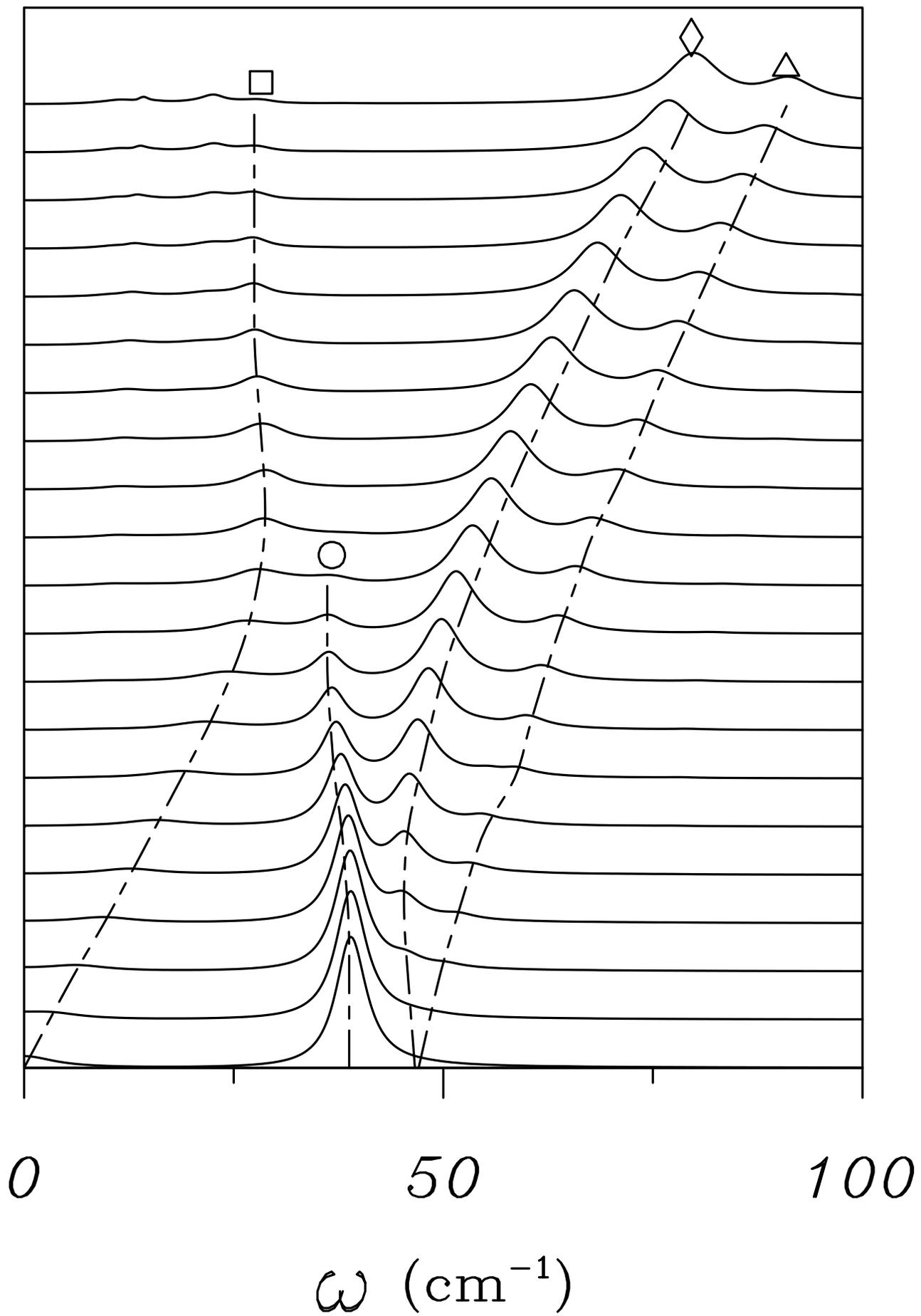
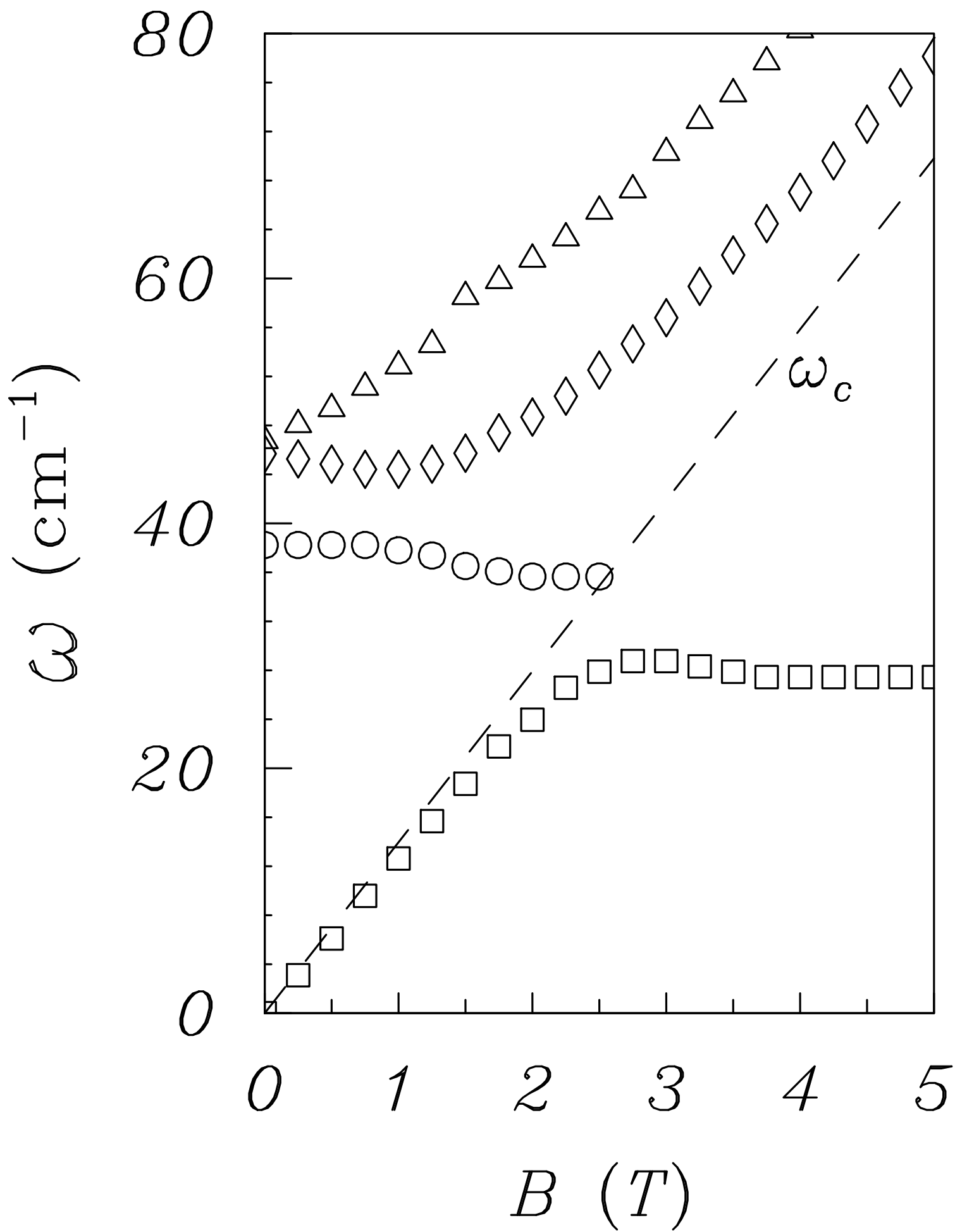
\end{document}